\documentstyle[fleqn,preprint,aps,floats,tighten,epsfig]{revtex}

\def\lsim{\mathrel{\rlap{\lower4pt\hbox{\hskip1pt$\sim$}}
    \raise1pt\hbox{$<$}}}         
\def\gsim{\mathrel{\rlap{\lower4pt\hbox{\hskip1pt$\sim$}}
    \raise1pt\hbox{$>$}}}         
\def\esim{\mathrel{\rlap{\raise2pt\hbox{$\sim$}}
    \lower1pt\hbox{$-$}}}         

\begin{document}

\begin{titlepage}

\title{\hfill {\rm\normalsize USITP/98-08} \\ \vskip-6pt
  \hfill {\rm\normalsize MPI-PhT/98-43} \\ \vskip-6pt
  \hfill {\rm\normalsize June 1998} \\~\\
  Clumpy Neutralino Dark Matter}

\author{Lars Bergstr\"om}
\address{Department of Physics, Stockholm University, Box 6730,
SE-113~85~Stockholm, Sweden;\\
lbe@physto.se }

\author{Joakim Edsj\"o}
\address{Center for Particle Astrophysics, University of California,\\
301 Le Conte Hall, Berkeley, CA 94720-7304, U.S.A;\\
edsjo@cfpa.berkeley.edu}

\author{Paolo Gondolo}
\address{Max Planck Institut f\"ur Physik, F\"ohringer Ring 6, 80805 
Munich, 
Germany;\\
gondolo@mppmu.mpg.de}

\author{Piero Ullio}
\address{Department of Physics, Stockholm University, Box 6730,
SE-113~85~Stockholm, Sweden;\\
piero@physto.se}

\maketitle

\begin{abstract}
   
We investigate the possibility to detect neutralino dark matter in a
scenario in which the galactic dark halo is clumpy.
 We find that under
customary assumptions on various astrophysical parameters,
the antiproton and continuum $\gamma$-ray signals from neutralino
annihilation in the  halo 
put the strongest limits on the clumpiness of a neutralino halo.
We argue that indirect detection 
through neutrinos from the Earth and the Sun should not be much
affected by clumpiness. We identify 
situations in parameter space where the $\gamma$-ray line,
positron and diffuse neutrino 
signals from annihilations in the halo may provide interesting
signals in upcoming 
detectors. 
\end{abstract}

\end{titlepage}


\section{Introduction}

The mystery of the dark matter in the Universe remains unsolved.  
Among the more plausible candidates (not only needed to solve the dark 
matter problem) can be found the neutralino, the lightest 
supersymmetric particle in the Minimal Supersymmetric Standard Model 
(MSSM) (for a review, see \cite{jkg}).  Another candidate is for 
instance the axion which is still a viable option for a narrow range 
of axion masses \cite{raffelt}.  Irrespective of the exact nature of 
the dark matter, there are reasons to believe that its distribution in 
the dark halos of galaxies need not be perfectly smooth 
\cite{SilkSzalay,SilkStebbins,kolb}.  For instance, early fluctuations 
in the dark matter may go non-linear long before photon decoupling, 
evading the argument of slow, linear growth after recombination.  
Also, if cosmic strings or other defects exist, they may seed the 
formation of density-enhanced dark matter clumps.

Since very little is known about the inherently non-linear problem of 
generating dark matter clumps, in this paper we will use a 
phenomenological approach where we simply assume the existence of
clumps 
with a given density profile, making up a certain fraction of the total 
mass of the Milky Way halo.  We investigate the effect of this clumpiness
on the various 
proposed detection methods for 
neutralino dark matter. Detection rates 
depend crucially on the neutralino distribution in
momentum and position space.
 Some detection rates, in particular those 
of antiprotons and photons generated by neutralino annihilations in 
the galactic halo, increase substantially compared to the case of a 
smooth dark matter distribution.  For a given set of parameters of the 
supersymmetric models (such as mass and couplings of the neutralinos) 
we can then use present experimental limits on these fluxes to bound 
the degree of clumpiness allowed in that particular dark matter model.  
Alternatively, given a positive experimental signature, we can 
identify regions in the combined parameter space of halo dark matter 
distribution and supersymmetric models to identify candidates 
consistent with the data.  This approach was used recently by three of 
us in connection with new data from the EGRET gamma ray detector 
\cite{BEU}.  Some of our results may be of interest also in the 
standard non-clumpy scenario, which is of course included in our 
treatment and is easily recovered by putting the fraction of the halo 
in the form of clumps equal to zero.


\section{Clumpiness in the Milky Way halo}
\label{sec:halo}

Present observational data give very poor constraints on the 
distribution of dark matter in the galaxy.  The dynamics of the outer 
satellites of the galaxy clearly indicates that luminous matter 
provides just a fraction of the total mass of the Milky Way and that 
the major contribution must come from a dark matter halo whose size is 
larger than the radius of the disk.  Nevertheless it is not possible 
to extract from present kinematic information any accurate knowledge 
of the density profile of the dark matter halo.  It is, however, 
natural to assume that galactic dark matter profiles obey a law of 
universality.  Then, a possible approach is to infer the functional 
form of the Milky Way halo density profile from the results of 
$N$-body simulations of hierarchical clustering in cold dark matter 
cosmologies, fitting the normalization parameters to known dynamical 
constraints.  This approach has been followed in Ref.~\cite{lpj}: 
among the general family of spherical density profiles,
\begin{equation}
  \rho(\vec{x}) = \rho_0\; \left(\frac{R_0}{|\vec{x}|}\right)^{\gamma} 
  \,\left[\frac{1+(R_0/a)^\alpha}{1+(|\vec{x}|/a)^\alpha}\right] 
  ^{(\beta-\gamma)/\alpha},
\label{eq:prof}
\end{equation}
it was considered the case of the Kravtsov et al.\ profile 
\cite{kravtsov} which is mildly singular towards the galactic centre 
with $\gamma \sim 0.2$--$0.4$, of the Navarro et al.\ profile 
\cite{navarro} which is more cuspy ($\gamma = 1$), and, for comparison 
the modified isothermal distribution, $(\alpha,\beta,\gamma)=(2,2,0)$, 
extensively used in dark matter detection computations.  

The dark matter density profile inferred in this way should be 
regarded as the function that describes the average distribution of 
dark matter in the galactic halo; the standard assumption which is 
generally made at this stage is that dark matter particles in the halo 
form a perfectly smooth `gas'.  This approach is in some way 
arbitrary: although the dark matter particle distribution has to be 
regarded as smooth on intermediate length scales, probably around 
0.01--1 kpc, there are reasons to question whether this is true on 
smaller scales.  We here entertain the possibility that at least a 
fraction of the dark matter in the halo is clustered in substructures 
with high matter density, `clumps' of dark matter.  Several authors 
have introduced clumpiness as a generic feature of cold dark matter 
cosmologies.  Silk and Stebbins \cite{SilkStebbins} have considered 
clump formation in cosmic string, texture and inflationary models, 
giving also predictions for survival to tidal disruption (see also 
Ref.~\cite{SilkSzalay}).  Kolb and Tkachev \cite{kolb} have studied 
isothermal fluctuations giving very high-density dark matter clumps.

Simulations of structure formation in the early Universe do not yet 
have the dynamical range to give predictions for the size and density 
distribution of small mass clumps (we focus here mainly on clumps of 
less than around $10^6$ solar masses which avoid the problem of 
unacceptably heating the disk~ \cite{SilkStebbins}).  The formation of 
clumps on all scales is however a generic feature of cold dark matter 
models which have power on all length scales.  If self-similarity is a 
guide, galaxy halos may form hierarchically in a similar way to that 
of cluster halos (see e.g.\ Ref.\ \cite{bepi}).

Rather than examining the different scenarios for clump 
formation, we take a more phenomenological approach and perform a
detailed discussion on the implications of clumpiness on neutralino 
dark matter searches.

We thus simply postulate that a fraction $f$ of the total dark matter is 
concentrated in clumps, which are assumed to be spherical bodies 
of typical mass $M_{cl}$ and matter density profile 
$\rho_{cl}(\vec{r_{cl}})$.
The total number of clumps inside the halo is given by:
\begin{equation} 
  N_c \sim \frac{f\,\cdot\,M_h}{M_{cl}}
\label{eq:numcl}
\end{equation} 
where $M_h$ is the total mass of the halo.  Two opposite scenarios 
seem to be plausible.  There might be few heavy clumps, with masses up 
to maybe $M_{cl} \sim 10^{6}$--$10^{8} M_\odot$ (above which the local 
gravitational distortion effects would be too severe); such massive 
bodies could in principle be identified from the analysis of the 
rotation curves of the galaxy, they may however have escaped 
observation so far because their detection in this way may be 
difficult if the fraction of dark matter in clumps is small, say 
$f\sim1\%$.  A second possibility is that clumps are much lighter, 
with $M_{cl}$ less than $10^{4}$--$10^{6} M_{\odot}$, in which case 
larger fractions of the halo mass might be in clumps (in the extreme 
scenario all of it).  In the many small clumps scenario, on which we 
mainly focus, we can define a probability density distribution of the 
clumps in the galaxy which in the limit of large $f$, to fulfill 
dynamical constraints, has to follow the mass distribution in the 
halo.

Consider a Cartesian coordinate system with origin in the galactic 
centre.  Then the probability for a given clump being in the volume 
element $d^{\,3}x$ at position $\vec{x}$ is
\begin{equation} 
  p_{cl}(\vec{x})\,d^{\,3}x = \frac{1}{M_h}\; \rho(\vec{x})\,d^{\,3}x
\label{eq:prob} 
\end{equation} 
which has the correct normalization $\int p_{cl} (\vec{x}) d^{\,3}x = 1$.

It is convenient to introduce the dimensionless parameter $\delta$
\begin{equation}
    \delta =\frac{1}{\rho_{0}} 
            \frac{\int d^{\,3}r_{cl}\,
                  \left(\rho_{cl}(\vec{r_{cl}})\right)^{2}}
                  {\int d^{\,3}r_{cl}\,\rho_{cl}(\vec{r_{cl}})}
\end{equation}
which gives the effective contrast between the dark matter density in 
clumps and the local halo density $\rho_0$.  For a dark matter density 
inside the clumps which is roughly constant, $\rho_{cl}$, it reduces 
to the form
\begin{equation}
  \delta = \frac{\rho_{cl}}{\rho_0}\;. 
\end{equation}
We show in Section~\ref{sec:gampbar} that in the many-clumps scenario 
it is just the product $f \delta$ which determines the increase of 
the signal compared to a smooth halo in most indirect detection methods.
The product $f \delta$ is directly related to the ratio of the total dark
mass in clumps to the volume of a typical clump.

\section{Supersymmetric models for dark matter}

Although it is possible that the halo dark matter may be composed of 
particles not yet predicted by particle physics models, it is very 
attractive to assume that they are weakly interacting massive 
particles (WIMPs).  Massive particles with weak interactions give a 
relic density which is of the right order of magnitude to explain the 
dark matter on all scales from dwarf galaxies and upwards.  We will 
consider a specific class of such particles, supersymmetric 
particles, which is general enough to illustrate the effects of 
clumpiness. Our results should be of more general validity, however.

\begin{table}
  \centering
  \begin{tabular}{lrrrrrrr}
  Parameter & $\mu$ & $M_{2}$ & $\tan \beta$ & $m_{A}$ & $m_{0}$ & 
  $A_{b}/m_{0}$ & $A_{t}/m_{0}$ \\
  Unit & GeV & GeV & 1 & GeV & GeV & 1 & 1 \\ \hline
  Min & -50000 & -50000 &  1.0 &     0 &   100 & -3 & -3 \\
  Max &  50000 &  50000 & 60.0 & 10000 & 30000 &  3 &  3 \\
  \end{tabular}
  \caption{The ranges of parameter values used in our scans of the 
  MSSM parameter space.  Note that several special scans aimed at 
  interesting regions of the parameter space have been performed.  In 
  total we have generated about 85000  models that are not 
  excluded by accelerator searches.}
  \label{tab:scans}
\end{table}

We work in the Minimal Supersymmetric Standard Model (MSSM) as defined 
in Refs.~\cite{haberkane,jkg}.  For details on our notation, see 
Ref.~\cite{coann}.  The lightest stable supersymmetric particle is in 
most models the neutralino, which is a superposition of the 
superpartners of the gauge and Higgs fields,
\begin{equation}
  \tilde{\chi}^0_1 = 
  N_{11} \tilde{B} + N_{12} \tilde{W}^3 + 
  N_{13} \tilde{H}^0_1 + N_{14} \tilde{H}^0_2.
\end{equation}
It is convenient to define the gaugino fraction of the lightest 
neutralino,
\begin{equation}
  Z_g = |N_{11}|^2 + |N_{12}|^2
\end{equation}
For the masses of the neutralinos and charginos we use the one-loop 
corrections as given in \cite{neuloop}.

The MSSM has many free parameters, but with some simplifying 
assumptions, we are left with 7 parameters, which we vary between 
generous bounds.  The ranges for the parameters are shown in 
Table~\ref{tab:scans}.  For the detection rates of neutralino dark 
matter we have used the rates as calculated in Refs.\ 
\cite{BEU,lpj,bg,beg,eg,joakim}.

We will throughout this paper assume that the 
neutralinos make up most of the dark matter in our galaxy.
We only consider therefore MSSM models which are cosmologically interesting, 
i.e.\ where the neutralinos can make up a major fraction of the dark 
matter in the Universe without overclosing it.  We will choose this 
range to be $0.025 < \Omega_{\chi}h^{2} <0.5$.  For the relic density 
calculations we have used the detailed calculations performed in 
Ref.\ \cite{coann}.

\section{Detection methods constraining clumpiness}
\label{sec:gampbar}

Some observational consequences of a clumpy dark matter halo have been 
pointed out previously, such as the obvious gain in gamma ray signal 
from annihilation in the halo since the flux from a particular volume 
element is proportional to the square of the dark matter density 
there~\cite{SilkStebbins,kolb,lake,bss,wass,bbm}.  Also, in Ref.\ 
\cite{pbar1} it was noted that the antiproton flux could be enhanced, 
although the treatment was sketchy and not entirely correct concerning 
the way the rescaling was done.  In Ref.\ \cite{mica}, it was 
investigated whether encounters with dark matter clumps on geophysical 
time scales could have left imprints in ancient mica.

As we show in this section, indirect detection through cosmic antiprotons
and gamma rays set the most stringent limits on clumpy neutralino dark
matter, therefore we investigate these cases first.

\subsection{Gamma-rays}

Since gamma rays produced in neutralino annihilations in the halo
travel in straight paths essentially without any absorption, and
since the annihilation rate and hence the flux would be enhanced by
clumps along a particular line-of-sight, the effects of clumpiness
are easy to understand.

Neutralino annihilation in the galactic halo may produce both a 
$\gamma$-ray flux with a continuum energy spectrum and monochromatic 
$\gamma$-ray lines.  

The continuum contribution (see Ref.~\cite{jkg} and references therein)
 is mainly due to the decay of $\pi^0$ mesons produced in jets
from neutralino annihilations.
To model the fragmentation process and extract 
information on the number and energy spectrum of the $\gamma$s 
produced we have used the Lund Monte Carlo {\sc Pythia} 
6.115~\cite{pythia}.  We have performed the simulation for 18 
neutralino masses between 10 and 5000 GeV and for the $c \bar{c}$, $b 
\bar{b}$, $t \bar{t}$, $W^{+} W^{-}$, $Z^{0} Z^{0}$ and $gg$ 
annihilation states.  For each final state and for each neutralino 
mass we have simulated $2.5 \times 10^5$ events which are tabulated 
logarithmically in energy.  For any given MSSM model, we then sum over 
the annihilation channels and interpolate in these tables.  For the 
annihilation channels not included in the simulations, like the ones 
with one gauge and one Higgs boson as well as those with two Higgs 
bosons the flux is calculated in terms of the flux from the simulated 
channels.  We include all two-body final states at the tree level (except 
light quarks and leptons) and the one-loop processes $Z \gamma$ 
and $gg$.  For final states with Higgs bosons, we let the Higgs bosons 
decay in flight by summing the contributions to the gamma flux from 
the Higgs decay products in the Higgs rest system and then boost the 
spectrum averaging over decay angles.  Given the annihilation 
branching ratios we then get the spectrum for any given MSSM model.  
The continuum signal lacks distinctive features and it might be 
difficult to discriminate from other possible sources.  It will 
however be a powerful tool to put constraints on the clumpiness 
parameters.

A much better signature than the continuum contribution is given by 
monochromatic $\gamma$-ray lines which arise from the loop-induced S-wave 
neutralino annihilations into the $2 \gamma$ and $Z \gamma$ 
final states and which have no conceivable background from known
astrophysical sources. The amplitude of these two processes in the MSSM
was computed only recently at full one loop level~\cite{2gamma,zgamma}.
Large deviations from previous partial results 
(see Ref.~\cite{jkg} and references therein) were found, 
in particular it was pointed out that a pure heavy Higgsino has a 
remarkably high annihilation branching ratio both into $2 \gamma$ and 
$Z \gamma$, adding at least a factor of 10 to previous estimates of the 
$2 \gamma$ line. A detailed phenomenological study is given in Ref.~\cite{lpj}
where a smooth halo scenario was considered and it was shown that the
monochromatic lines could be detected by the new generation of space- and
ground-based $\gamma$-ray experiments, provided that a sensible enhancement
of the dark matter density is present towards the galactic centre.
We examine here the perspectives of detecting the continuum and the line
signals in a given clumpy scenario.

Consider a detector with an angular acceptance $\Delta\Omega$ pointing 
in a direction of galactic longitude and latitude $(\ell, b)$.  The 
gamma ray flux from neutralino annihilations at a given energy $E$ is 
given by
\begin{equation}
  \Phi_\gamma (E,\,\Delta \Omega,\, \ell,\,b) 
  \simeq 1.87 \cdot 10^{-8} 
  \, \frac{d{\cal S}}{dE} \,
  \langle\,J\left(\ell,\,b \right)\,\rangle\,(\Delta\Omega)
  \;\;\rm{cm}^{-2}\;\rm{s}^{-1}\;\rm{sr}^{-1}\;. 
  \label{eq:flux}
\end{equation}
In this formula we have defined a factor $d{\cal S}/dE$ which depends on 
the nature of relic WIMPs. For the continuum $\gamma$-ray signal, 
the $2\gamma$ line and the $Z\gamma$ line signal, respectively, it
is given by: 
\begin{eqnarray}
  \left(\frac{d{\cal S}}{dE}\right)_{{\rm cont.}~\gamma} & 
  \simeq &\left( \frac{10\,\rm{GeV}}{M_\chi}\right)^2
  \cdot\sum_F \left( \frac{v\sigma_F}{10^{-26}\ {\rm cm}^3\; 
   {\rm s}^{-1}}\right)\,\frac{dN_\gamma^{\,F}}{dE} \nonumber \\
  \left(\frac{d{\cal S}}{dE}\right)_{2\gamma} & \simeq & 
  \left( \frac{10\,\rm{GeV}}{M_\chi}\right)^2
  \,\left( \frac{2\;v\sigma_{2\gamma}}{10^{-26}\ {\rm cm}^3\; 
   {\rm s}^{-1}}\right)\, \delta\left(E - M_\chi\right) \nonumber \\
  \left(\frac{d{\cal S}}{dE}\right)_{Z\gamma} & 
  \simeq &\left( \frac{10\,\rm{GeV}}{M_\chi}\right)^2
  \,\left( \frac{v\sigma_{Z\gamma}}{10^{-26}\ {\rm cm}^3\; 
   {\rm s}^{-1}}\right)\, \delta\left(E - M_\chi\,\left(1 - \frac{{M_Z}^2}
  {4\,{M_\chi}^2}\right)\right).
\end{eqnarray}
Here $M_\chi$ is the neutralino mass, $F$ are the allowed final states 
which contribute to the continuum signal as specified above.  For each 
of these, $v\sigma_F$ is the annihilation rate and 
$dN_\gamma^{\,F}/dE$ is the differential energy distribution of 
produced photons.  The product of relative velocity and cross section 
$v\sigma_{2\gamma}$ is the annihilation rate into the $2\gamma$ final 
state (as given in Ref.~\cite{2gamma}).  Similarly, 
$v\sigma_{Z\gamma}$ is the rate into the $Z\gamma$ final state (as 
given in Ref.~\cite{zgamma}).  In Eq.~(\ref{eq:flux}) the dependence 
of the flux on the dark matter distribution, the direction of 
observation $(\ell,b)$
and the angular acceptance of the detector $\Delta\Omega$ is contained in 
the factor $\langle\,J\left(\ell,\,b 
\right)\,\rangle\,(\Delta\Omega)$.  If we assume a spherical dark 
matter halo in the form of a perfectly smooth distribution of 
neutralinos, it is equal to
\begin{eqnarray}
\langle\,J\left(\psi\right)\,\rangle\,(\Delta\Omega) =
\frac{1} {8.5\, \rm{kpc}}\,\frac{1} {\Delta\Omega}
\;\int_{\Delta\Omega}\,d\Omega'\; 
\int_{line\;of\;sight}\,dL\;
\left(\frac{\rho(L,\,\psi')}
{0.3\,{\rm GeV}/{\rm cm}^3}\right)^2\;.
\label{eq:jpsi}
\end{eqnarray}
Here $L$ is the distance from the detector along the line of sight,
$\psi$ is the angle between the direction of observation and that of the 
galactic center, related to $(\ell,\,b)$ by $\cos\psi = \cos\ell \, \cos b$.
The integration over $d\Omega'$ is performed over the solid angle 
$\Delta\Omega$ centered on $\psi$. 

\begin{figure}
\centerline{\epsfig{file=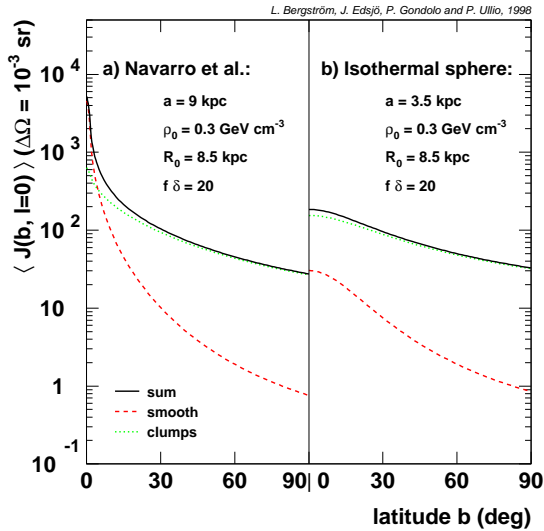,width=0.49\textwidth}}
\caption{The value of $\langle\,J(\psi)\,)\rangle\,(\Delta\Omega)$ for 
two different halo profiles. The contribution from the smooth and 
clumpy component are also given.}
\label{fig:jpsi}
\end{figure}

We shall now examine the consequences of introducing clumps in the 
halo.  The continuum $\gamma$-ray signal in the few clumps scenario 
has as mentioned been examined in some detail in the literature.  
Following the approach of Refs.~\cite{lake,bss} and estimating the most 
probable distance for the nearest clump, we find in terms of the 
quantities introduced above that $\langle\,J\,\rangle$ in the 
direction of the nearest clump is of the order
\begin{equation}
  \langle\,J\left(\psi_{cl}\right)\,\rangle\,(\Delta\Omega) 
  \gsim \left(\frac{4\,\pi}{3}\right)^{2/3}\,
  \frac{\delta}{8.5\,\rm{kpc}}\,\frac{1}{\Delta\Omega}
  \left(f^2\,\frac{M_{cl}}{\rho_0}\right)^{1/3}\;,
\label{eq:1clump}
\end{equation}
where the density profile inside the clump was considered roughly constant
and the angular acceptance of the detector $\Delta\Omega$ was supposed
to be greater or equal to the field of view of the clump
$\Delta\Omega_{cl} \sim (f/\delta)^{2/3}$.
We consider as an example the same choice of parameters as in 
Ref.~\cite{bss}: 
$f \sim 0.01$, $M_{cl} \sim 10^{8} M_\odot$, $\delta \sim 10^3$. In 
this case we find $\Delta\Omega_{cl} \sim 4 \cdot 10^{-4}$~sr, and taking 
$\Delta\Omega = 10^{-3}$~sr, we obtain 
$ \langle\,J\left(\psi_{cl}\right)\,\rangle\,\sim 3\cdot 10^4$ which 
we can compare with the analogous quantity in a smooth halo scenario. 
In Fig.~7 of Ref.~\cite{lpj} the values of 
$\langle\,J\,\rangle\,(\Delta\Omega = 10^{-3} \rm{~sr})$ for a 
detector with $\Delta\Omega = 10^{-3} \rm{~sr}$
pointing towards the galactic centre are displayed (see also our
Fig.~\ref{fig:jpsi}); the  
value of $\langle\,J\left(\psi_{cl}\right)\,\rangle$ is about one 
half of the value for the most singular Navarro et al. profile,
which gives detectable $\gamma$-ray lines for a relevant portion of
MSSM models.

The clump in our example might therefore be a very bright dark matter 
source, and a signal from neutralino annihilation into monochromatic 
photons in its direction could potentially be detected with an Air 
Cherenkov Telescope (ACT)\@.  In practice, the probability is small of 
detecting such a signal randomly pointing an instrument with a small 
angular acceptance.  It might be of some help to combine ground- and 
satellite-based observations.  A satellite detector, which has a wide 
field of view but also a much smaller effective area with respect to 
an ACT, may search in the continuum $\gamma$-ray spectrum for brighter 
spots in the sky which have no luminous counterpart.  Such signals 
might then be identified as clumps of dark matter if one would detect 
with an ACT the $\gamma$-ray lines from neutralino annihilations.  For 
such a method to be practical, higher overdensities $\delta$ may be 
needed.  It should also be kept in mind that Eq.~(\ref{eq:1clump}) 
gives just a qualitative feature of the possible result; the 
possibility for the nearest clump of being much further away or a more 
realistic density profile may change that result by orders of magnitude.

Much firmer predictions may be formulated in the many small clumps 
scenario; in this case we assume that most of the clumps cannot be 
resolved even by a detector with a rather small angular acceptance, 
say about $\Delta\Omega \sim 10^{-3}$ sr.  There might still be some 
clumps which are resolvable just because they happen to be nearby and 
these should be treated as in the previous case.

{}From Eq.~(\ref{eq:prob}), the probability for a clump of being at a 
line of sight distance $(L,\,L+dL)$, a viewing angle defined by 
$(\cos\psi,\,\cos\psi+d\cos\psi)$ and at some azimuthal angle with 
respect to the direction of the galactic center $(\phi,\,\phi+d \phi)$ 
is given by
\begin{equation} 
  p_{cl}(L,\psi)\;dL\,d\cos \psi\,d\phi = \frac{1}{M_h}\; 
  \rho(L,\psi)\,L^2\;dL\,d\cos \psi\,d\phi\;.
\label{eq:prob2} 
\end{equation}
Assuming that the clumps can be regarded as point-like sources, we can 
derive the analogue of Eq.~(\ref{eq:jpsi}) (as in the latter we factorize out 
1/4$\pi$): 
\begin{eqnarray}
  \langle\,J(\psi)\,\rangle\,(\Delta\Omega) & = &
  \frac{1} {8.5\,\rm{kpc}}\,
  \frac{N_{cl}} {\Delta\Omega}
  \;\int_{\Delta\Omega}\,d\Omega'\; 
  \int_{line\;of\;sight}\,dL\;p_{cl}(L,\psi') \cdot \nonumber \\
  && \cdot \frac{1}{L^2}\,\int d^{\,3}r_{cl}\,
  \left(\frac{\rho_{cl}(\vec{r_{cl}})}{0.3\,{\rm GeV}/{\rm cm}^3}
  \right)^2  .
\end{eqnarray}
Taking  Eqs.~(\ref{eq:numcl}) and (\ref{eq:prob2}) into account, this can
be rewritten as
\begin{eqnarray}
  \langle\,J(\psi)\,\rangle\,(\Delta\Omega) &\sim& 
  \frac{1} {8.5\,\rm{kpc}}
  \,\frac{f \delta} {\Delta\Omega}
  \left(\frac{\rho_{0}}{0.3\,{\rm GeV}/{\rm cm}^{3}}\right) \cdot \nonumber \\
  && \cdot\int_{\Delta\Omega}\,d\Omega'\; 
  \int_{line\;of\;sight}\,dL\;
  \left(\frac{\rho(L,\psi')}
  {0.3\,{\rm GeV}/{\rm cm}^3}\right)\;.
  \label{eq:jpsicl}
\end{eqnarray}

Comparing Eq.~(\ref{eq:jpsicl}) with Eq.~(\ref{eq:jpsi}), we realize 
that in a scenario with many small clumps the angular dependence of 
the signal is different from the one in the smooth halo scenario. This is 
illustrated in Fig.~\ref{fig:jpsi} for two different halo models,
picking as an example $f\delta = 20$:
 a 
Navarro et al.\  profile (Eq.~(\ref{eq:prof}) with 
$(\alpha,\beta,\gamma)=(1,2,1)$ and, in our example, $\rho_0 = 
0.3$~GeV/cm$^3$, $a = 9$~kpc) and a modified isothermal sphere 
(Eq.~(\ref{eq:prof}) with 
$(\alpha,\beta,\gamma)=(2,2,0)$, $\rho_0 = 0.3$~GeV/cm$^3$, $a = 
3.5$~kpc).  The parameter 
$f\delta$ mainly determines the relative importance of the smooth and 
clumpy components.  An interesting feature, shown in the Figure for 
the Navarro et al.\ profile, is a possible break in the angular 
spectrum.  This could be a possible signature to discriminate the 
signal from neutralino annihilations into continuum $\gamma$-rays from 
the galactic $\gamma$-ray background, and may be indeed suggested by a 
recent analysis of EGRET data~\cite{BEU}.

\begin{figure}
\centerline{\epsfig{file=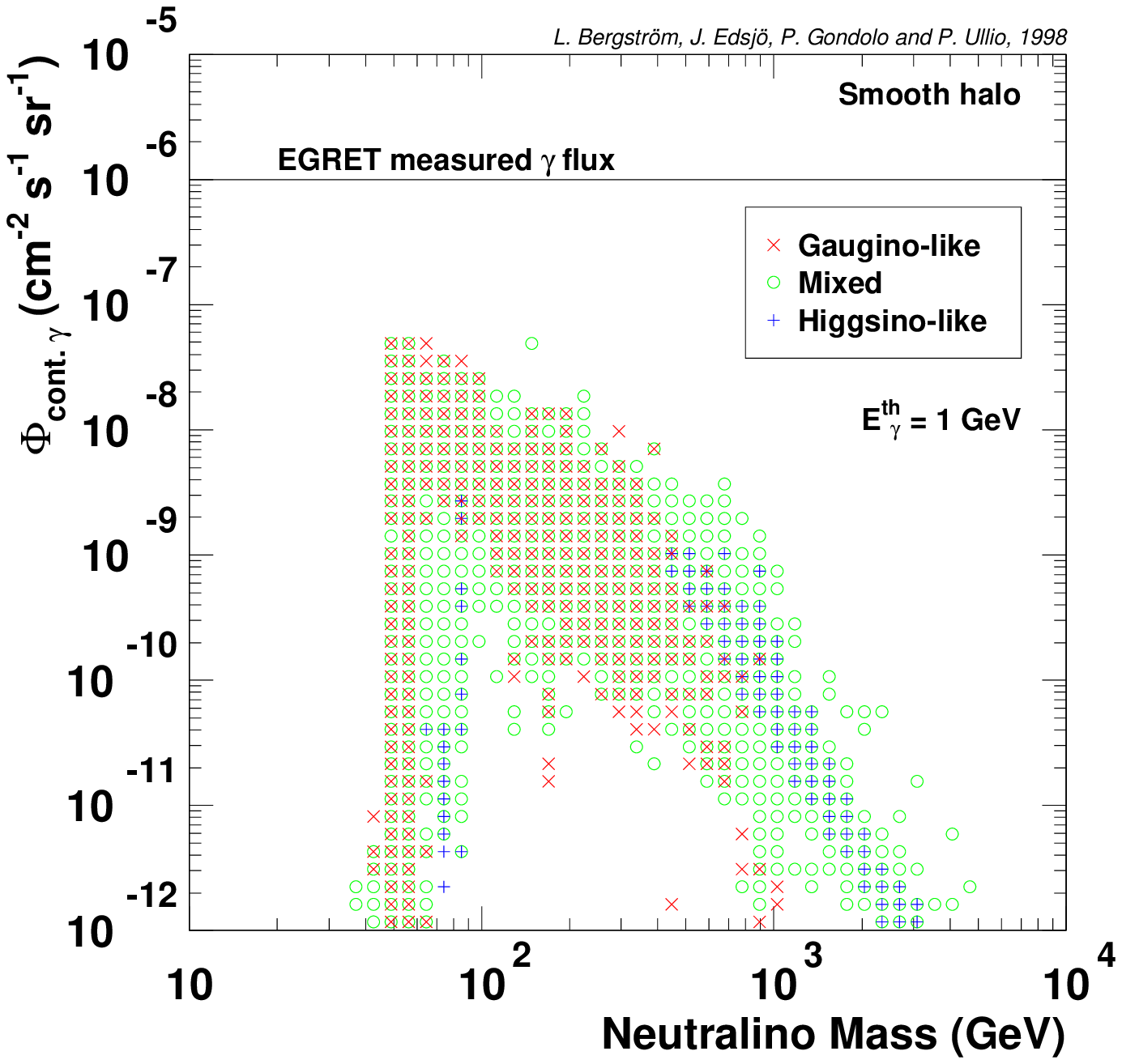,width=0.49\textwidth}
\epsfig{file=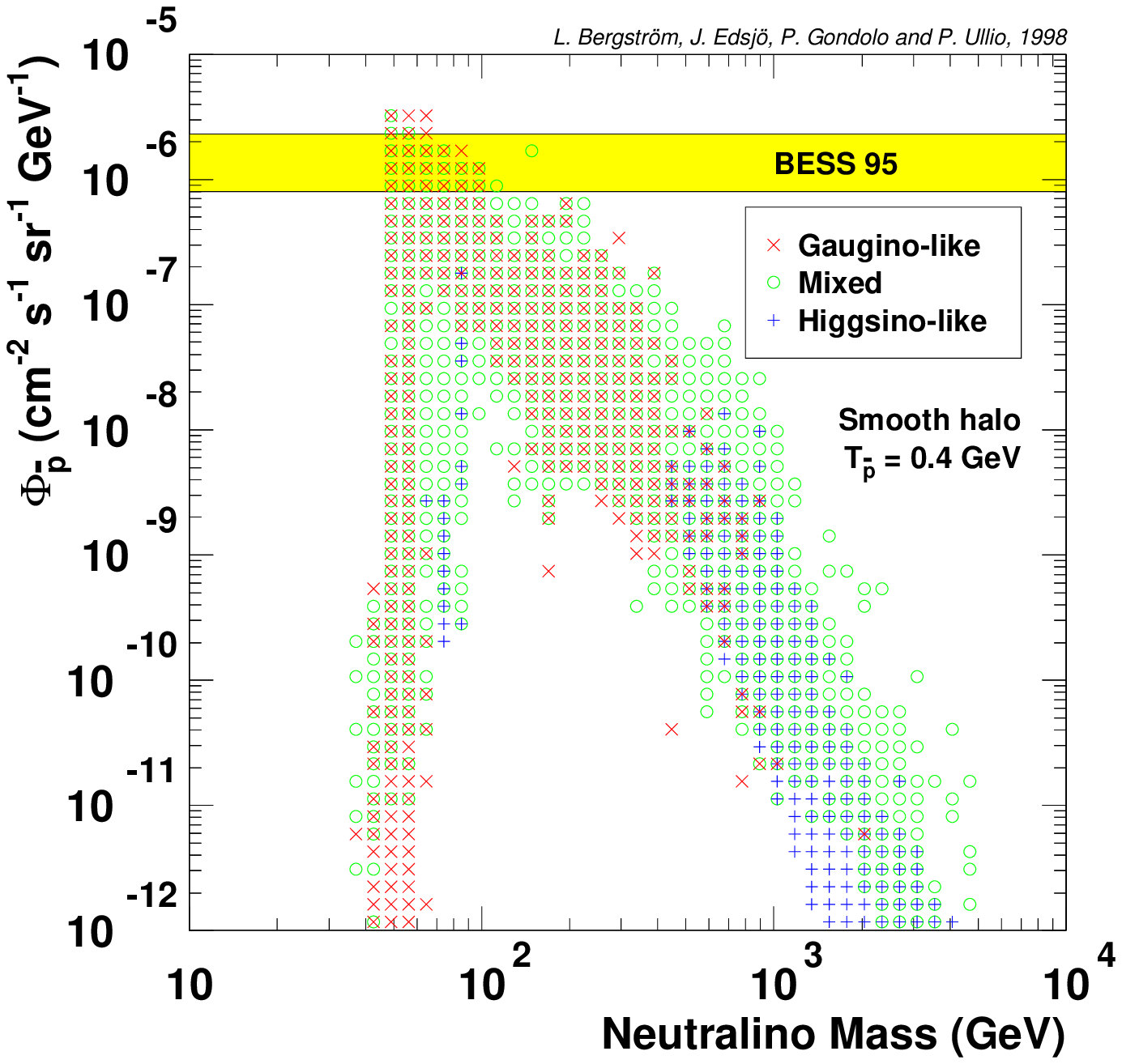,width=0.49\textwidth}}
\caption{The signal of (a) continuum gamma and (b) antiprotons versus 
the neutralino mass. Only models with $0.025 < \Omega_\chi h^2 <0.5$ have 
been included in this and the following figures.}
\label{fig:pgacpbar}
\end{figure}

We are now ready to give predictions for the $\gamma$-ray flux from 
neutralino annihilations.  To minimize the impact of the halo model 
and of experimental uncertainties, we concentrate on the flux at high 
latitudes, $b > 60^\circ$ and $0^\circ < \ell < 360^\circ$ 
($\Delta\Omega = 0.84$~sr), rather than considering the flux towards 
the galactic centre which as shown in Fig.~\ref{fig:jpsi} is 
maximal.  The modified isothermal profile of Fig.~\ref{fig:jpsi} 
gives
\begin{equation}
        \langle\,J(90^\circ)\,\rangle^{smooth}\,(0.84 sr) +
        \langle\,J(90^\circ)\,\rangle^{clumps}\,(0.84 sr) \simeq
        0.93 \cdot (1 + 1.8 \cdot f\delta) \; .
\label{eq:gamresc}
\end{equation}
For simplicity we have made the reasonable assumption that $f$ is 
small.  If that is not true we have to replace $1$ by $(1-f)^2$ in the 
above equation (as well as Eq.~(\ref{eq:pbarresc}) below).  The 
analogous estimates with any of the halo models considered in 
Ref.~\cite{lpj} are within a factor of 2 of the value given in 
Eq.~(\ref{eq:gamresc}).  There is therefore a very weak halo model 
dependence in these results.  In Fig.~\ref{fig:pgacpbar} (a) we plot 
the integrated $\gamma$-ray flux above the energy threshold $E_{th} = 
1$~GeV for our set of MSSM models in the smooth halo scenario.  Also 
shown in the figure is the corresponding $\gamma$-ray flux measured by 
the Energetic Gamma Ray Experiment Telescope (EGRET) as inferred from 
the analysis in Ref.~\cite{sreekumar}:
\begin{equation}
        \Phi_{\gamma}(E>1~\rm{GeV}) = (1.0 \pm 0.2) \times 10^{-6}
        ~\mbox{cm$^{-2}$ s$^{-1}$ sr$^{-1}$}\;.
        \label{eq:gambkg}
\end{equation}
We can compare with this value to obtain a constraint on the allowed 
values of the parameter $f\delta$.  It is however useful to analyse 
this together with the analogous constraint we can derive in the 
scenario of many small clumps from neutralino annihilations into 
cosmic ray antiprotons.

\subsection{Antiprotons}

Neutralino annihilations of relic neutralinos in the galaxy may 
produce cosmic ray antiprotons (\cite{jkg} and references 
therein,~\cite{pbar1,pbar2}) mainly from jets, in a process which is 
analogous to the case of continuum $\gamma$-rays.  To model the 
fragmentation process and extract information on the number and energy 
spectrum of the antiprotons produced we have again used the Lund Monte 
Carlo {\sc Pythia} 6.115 and applied the same tabulation technique as 
for the production of photons.  Including the same set of final states 
and treating the Higgs bosons in the same way, for any given MSSM 
model we can then obtain the energy spectrum of antiproton dark matter 
sources.

If we assume a smooth distribution of WIMPs in the galaxy, the production
rate of antiprotons in the volume element $d^{\,3}x$ at the galactic 
position $\vec{x}$ is given by
\begin{equation}
  \frac{d{\cal R}_{sm}(\vec{x})}{dT}\,d^{\,3}x =  
  \left( \frac{\rho(\vec{x})}{M_\chi}\right)^2
  \cdot\sum_F v\sigma_F\,\frac{dN_{\bar{p}}^{\,F}}{dT}\;d^{\,3}x
\label{eq:pbarsm}
\end{equation}
where $T$ is the kinetic energy of the antiprotons. We will not discuss here 
the few-clumps scenario as those predictions are extremely sensitive
to the parameters which define the model.
We focus instead on the many small clumps scenario, treating again single
clumps as point-like sources. In this case we find: 
\begin{eqnarray}
  \frac{d{\cal R}_{cl}(\vec{x})}{dT}\,d^{\,3}x &=&  
  N_{cl}\,p_{cl}(\vec{x})\;\int d^{\,3}r_{cl}\,
  \rho_{cl}^{2}(\vec{r_{cl}})\;  
  \cdot\sum_F v\sigma_F\,\frac{dN_{\bar{p}}^{\,F}}{dT}\;d^{\,3}x  \nonumber \\
  &=& f\,\delta\;\frac{\rho_0\;\rho(\vec{x})}{{M_\chi}^2}
  \cdot\sum_F v\sigma_F\,\frac{dN_{\bar{p}}^{\,F}}{dT}\;d^{\,3}x\;.
\label{eq:pbarcl}
\end{eqnarray}

It is not straightforward to simulate the propagation of charged particles
in the galaxy. Different models have been proposed, and no consensus has 
been established yet. We present results in the limit in which the 
propagation is modeled by pure diffusion, using the analytic solution
derived in Ref.\ \cite{pbar2} to which we refer for further 
details. 

\begin{figure}
\centerline{\epsfig{file=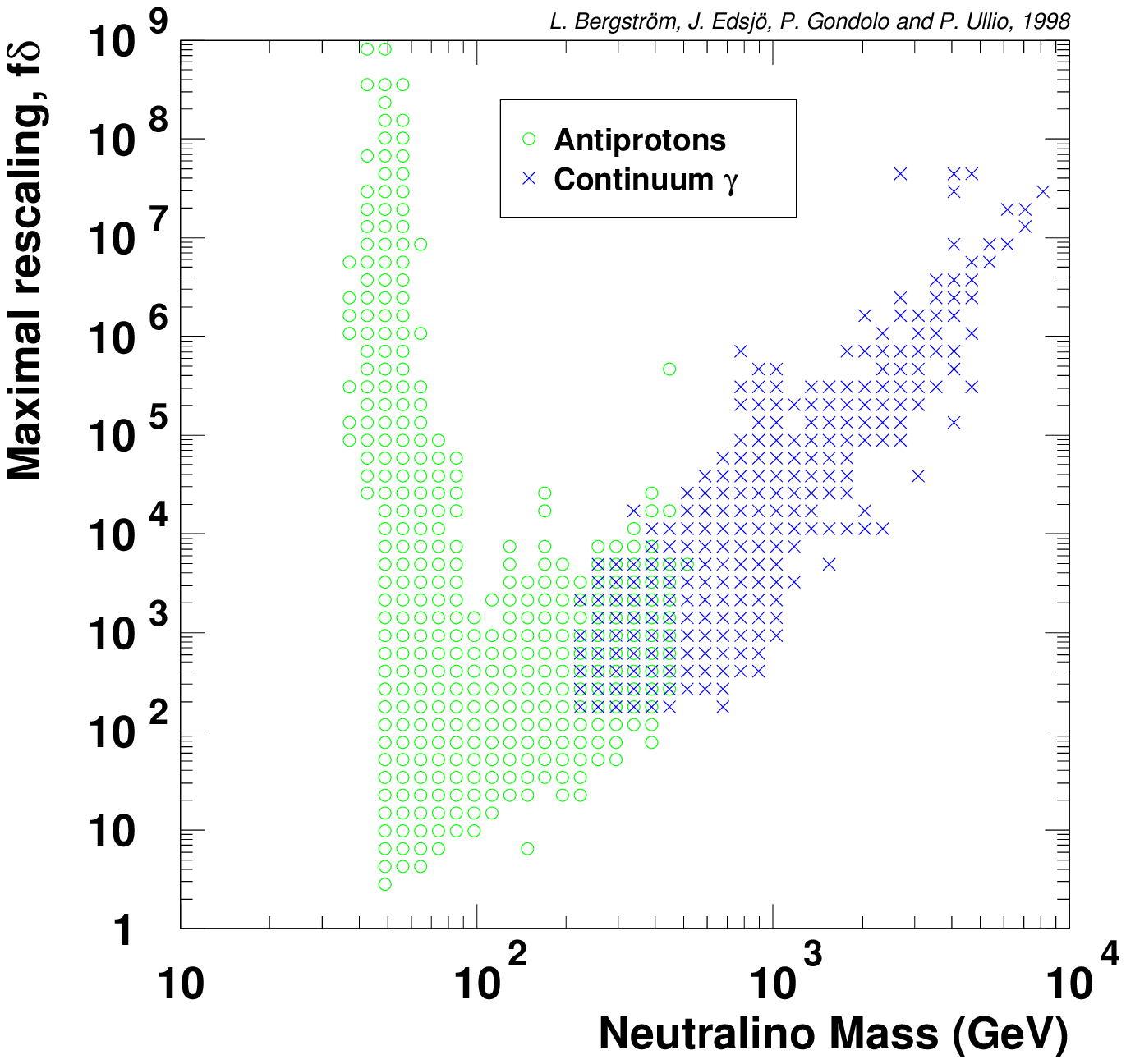,width=0.49\textwidth}
\epsfig{file=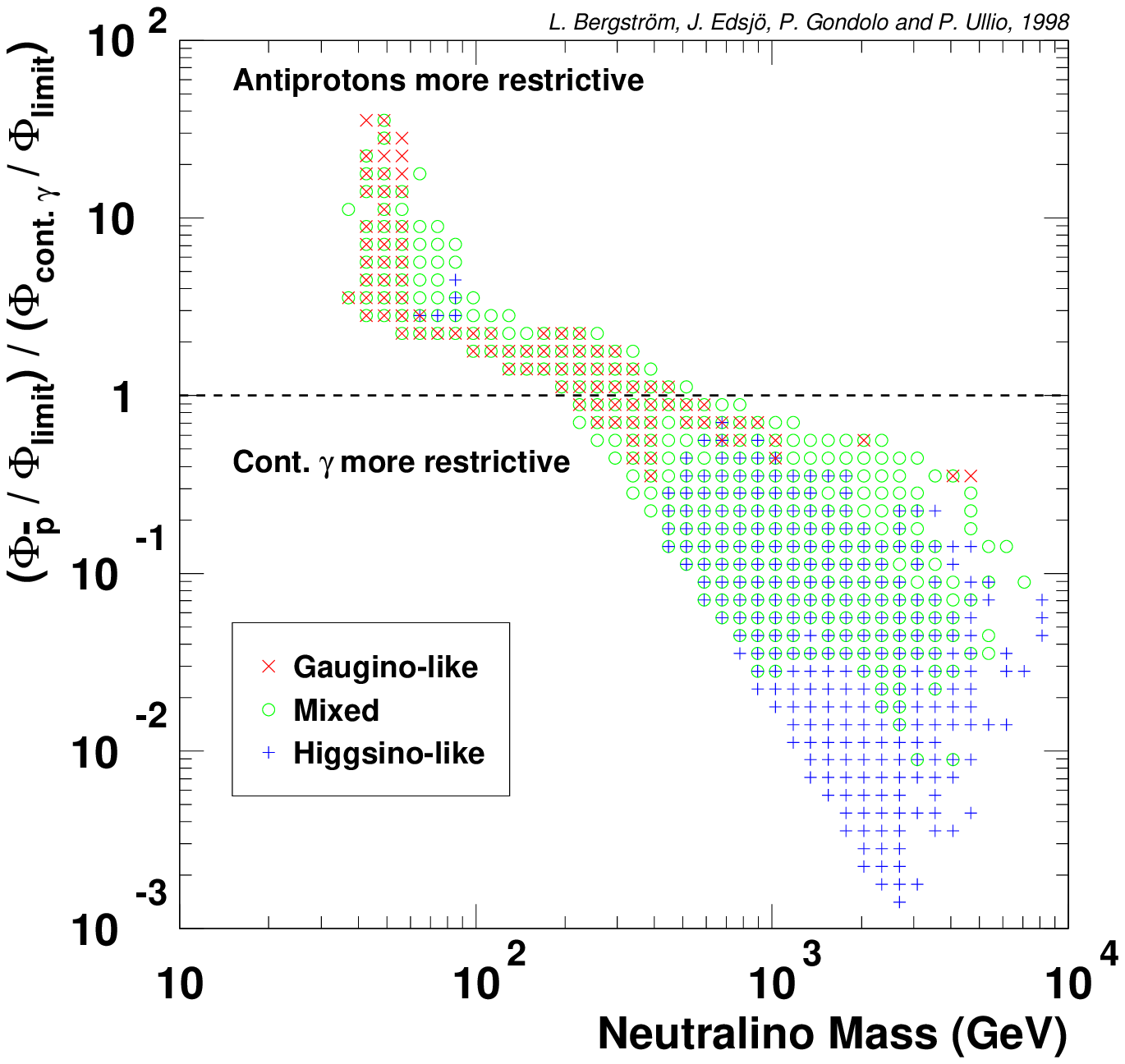,width=0.49\textwidth}}
\caption{The maximal rescaling allowed by the present limits on the 
antiproton flux and the continuum gamma ray flux. }
\label{fig:resc}
\end{figure}

The BESS collaboration \cite{bess} has in a recent measurement of 
cosmic ray antiprotons found that the spectrum for the antiproton flux 
versus the kinetic energy $T$ is consistent with being flat for $T$ in 
the range between 180 MeV and 1.4 GeV. We consider the value of the 
measured flux at some low value of the kinetic energy, where the 
`trivial' antiproton flux generated by cosmic-ray reactions in the 
interstellar medium is believed to be less dominant.  At $T = 400$ MeV 
the result found by BESS is
\begin{equation}
        \Phi_{\bar{p}}(T=400~\rm{MeV}) =
        1.4^{+0.9}_{-0.6} \times 10^{-6}~\bar{p}
        ~\mbox{cm$^{-2}$ s$^{-1}$ sr$^{-1}$ GeV$^{-1}$}\;.
\label{eq:bess400}
\end{equation}
In Fig.~\ref{fig:pgacpbar} (b) we compare this value with the 
predictions for antiprotons from neutralino annihilations in a smooth 
halo scenario (i.e.\ the source given as in Eq.~(\ref{eq:pbarsm})) at 
the same energy, using for the diffusion model the same set of 
parameters as in Ref.~\cite{pbar2}, considering appropriate values of 
the solar modulation parameters and picking as halo profile the 
modified isothermal distribution.  It is indeed tempting to conclude 
that some of our models are already excluded by the BESS measurement.  
However, one has to keep in mind the big uncertainties involved, 
mainly in the antiproton propagation; for instance it is not clear how 
large a fraction of antiprotons generated in the halo can penetrate 
the wind of cosmic rays leaving the disk~\cite{ptuskin}.  We introduce 
in the flux predictions a rescaling factor $k$ which contains the 
uncertainties deriving from the choice of the parameters which define 
the propagation model considered and from possible deviations from 
this simple approach.

We consider now the many small clumps scenario. The production rate of 
antiprotons in this case is given by Eq.~(\ref{eq:pbarsm}); the strength
of the signal compared to the smooth case is again mainly determined
by the product $f\delta$. At $T = 400$ MeV and for the same halo profile
considered above, we find:
\begin{equation}
        \Phi_{\bar{p}} =
        k (1+ 0.75 \cdot f\delta) \cdot \Phi_{\bar{p}}^{smooth}.
\label{eq:pbarresc}
\end{equation}
We have checked that the coefficient $0.75$ depends very weakly on the 
halo profile considered and on $T$.  A conservative limit on the 
clumpiness parameter $f\delta$ can be obtained choosing the 
uncertainty factor $k$ as
\begin{equation}
        k \in [0.2, 5]\;.
\label{eq:kint}
\end{equation}
We consider a value of $f\delta$ excluded if the whole range of possible 
antiproton fluxes given by Eqs.~(\ref{eq:pbarresc}) and (\ref{eq:kint}) 
exceeds the measured value, Eq.~(\ref{eq:bess400}).

\begin{figure}
\centerline{\epsfig{file=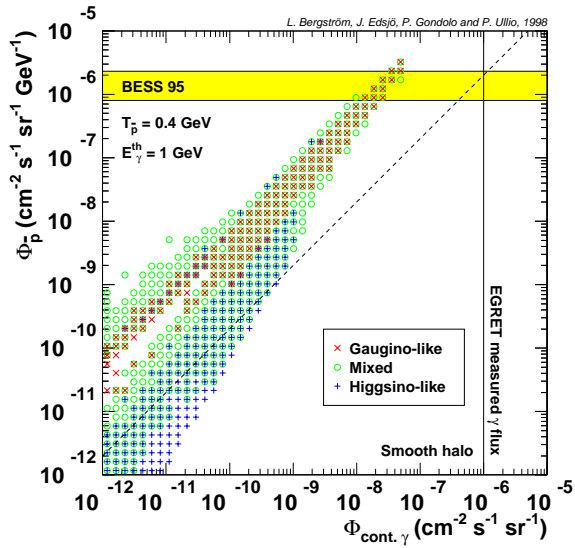,width=0.49\textwidth}}
\caption{The antiproton flux versus the continuum $\gamma$ flux for a 
smooth halo.}
\label{fig:pbarvspgac}
\end{figure}

\subsection{Determining the clumpiness factor $f\delta$ }

We have shown that in the many small clumps scenario the signals from 
dark matter annihilations into $\gamma$-rays and antiprotons depend 
critically on the clumpiness parameter $f\delta$.  Focusing on the 
MSSM, we use the rescalings derived in Eqs.~(\ref{eq:gamresc}) and 
(\ref{eq:pbarresc}) to determine for each supersymmetric model the 
maximal value of $f\delta$ for which the experimental constraints on 
the fluxes of continuum photons and antiprotons are not violated.  
This is shown in Fig.~\ref{fig:resc} (a), where the maximal rescaling 
is given versus the neutralino mass, and where we use different 
symbols to indicate which of the two bounds is more restrictive.  As 
can be seen, the antiproton flux puts the highest constraints on the 
clumpiness at low masses, whereas the continuum gammas put better 
constraints at higher masses.  We see that the present experimental 
limits constrain $f\delta \lsim 10^{9}$ for all masses.  

As shown in Fig.~\ref{fig:pbarvspgac}, the two signals are strongly 
correlated since they are both produced from jets.  In this sense the 
information we get from the two experimental limits is not entirely 
complementary.  At higher masses, both fluxes go down since they are 
both proportional to $1/M_{\chi}^{2}$, but the correlation also 
decreases since the antiproton fluxes are only given in a small energy 
interval while the gamma ray fluxes are integrated above a threshold.  
Hence the antiproton flux in a given low energy interval decreases 
more than the gamma ray flux as we go to higher neutralino masses.  In 
Fig.~\ref{fig:resc} (b) we analyse how restrictive one detection 
method is compared to the other.

Having derived for each of the MSSM models in our sample the maximal 
allowed value for the clumpiness parameter $f\delta$, in the next 
section we analyse the consequences of this result for other indirect 
detection methods of neutralino dark matter.

\section{Other detection methods}

In this section we consider the many small clumps scenario with the 
highest possible value of $f\delta$ as given in the previous section 
and investigate what effect that has on other dark matter searches.
We fix again as smooth halo distribution to compare with 
the modified isothermal distribution, Eq.~(\ref{eq:prof}) with 
$(\alpha,\beta,\gamma)=(2,2,0)$, $\rho_{0}=0.3$ GeV cm$^{-3}$, $a=3.5$ kpc 
and $R_{0}=8.5$ kpc.

\subsection{Monochromatic $\gamma$-ray lines}

As we have seen, the same scaling applies to the continuum and the 
line $\gamma$-ray signals, it is therefore straightforward to derive 
the maximal fluxes of monochromatic photons from neutralino 
annihilations.

We perform this analysis in light of the potential of the next generation 
of satellite-based $\gamma$-ray detectors, and in particular of the proposed
Gamma-ray Large Area Space Telescope (GLAST)~\cite{glast}. 
To prevent uncertainties due to the 
choice of the dark matter halo profile to play any role in the following 
discussion, we fix again as field of view a 0.84~sr cone in the direction
$b = 90^\circ$. In the actual experiment the detector will collect data with
a $4 \pi$~sr angular acceptance; as for most halo profiles the ratio signal 
to squared root of the background is greatly enhanced towards the galactic 
centre, the predictions we show are an underestimate of the possible
results.  

\begin{figure}
\centerline{\epsfig{file=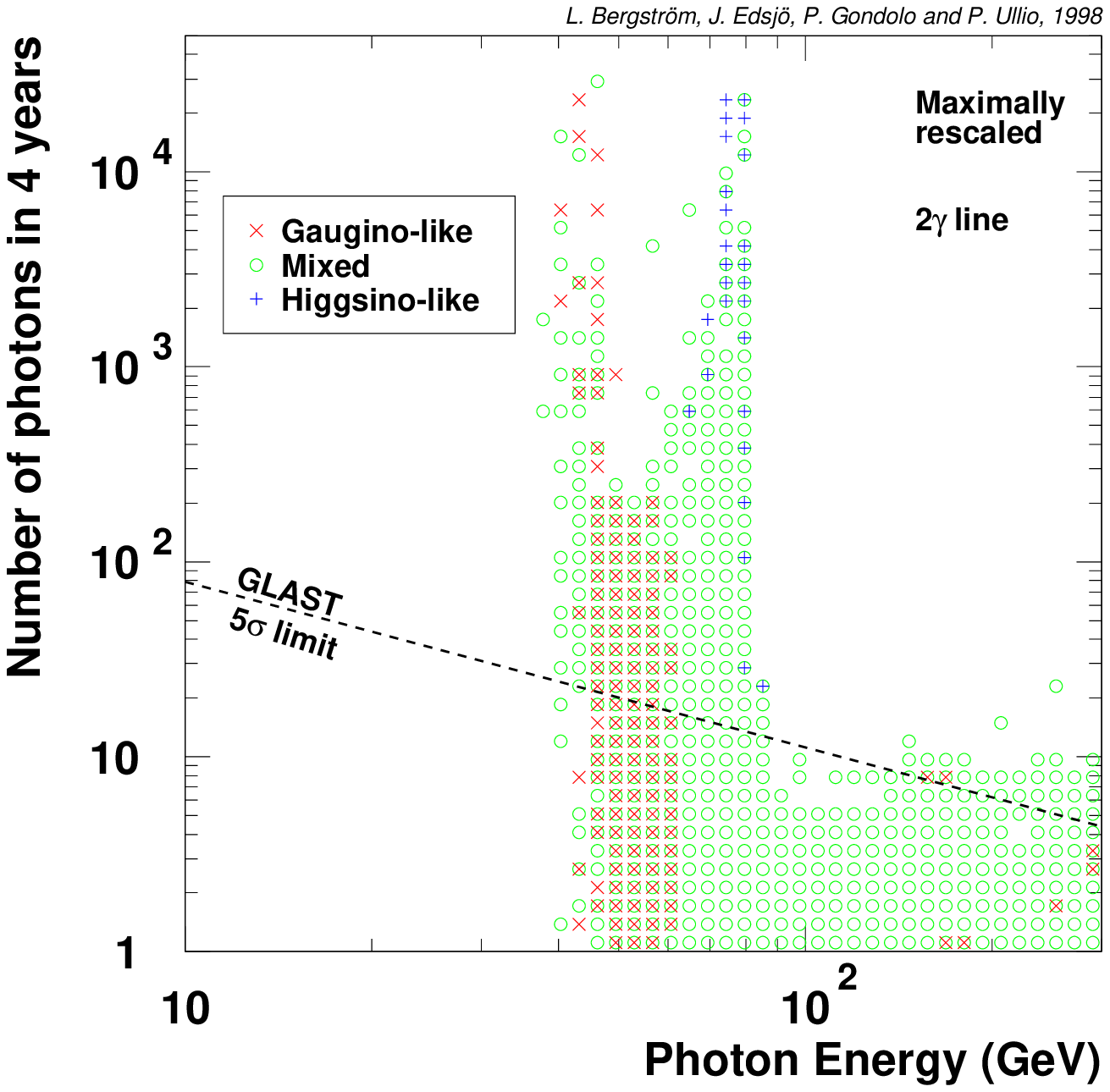,width=0.49\textwidth}
\epsfig{file=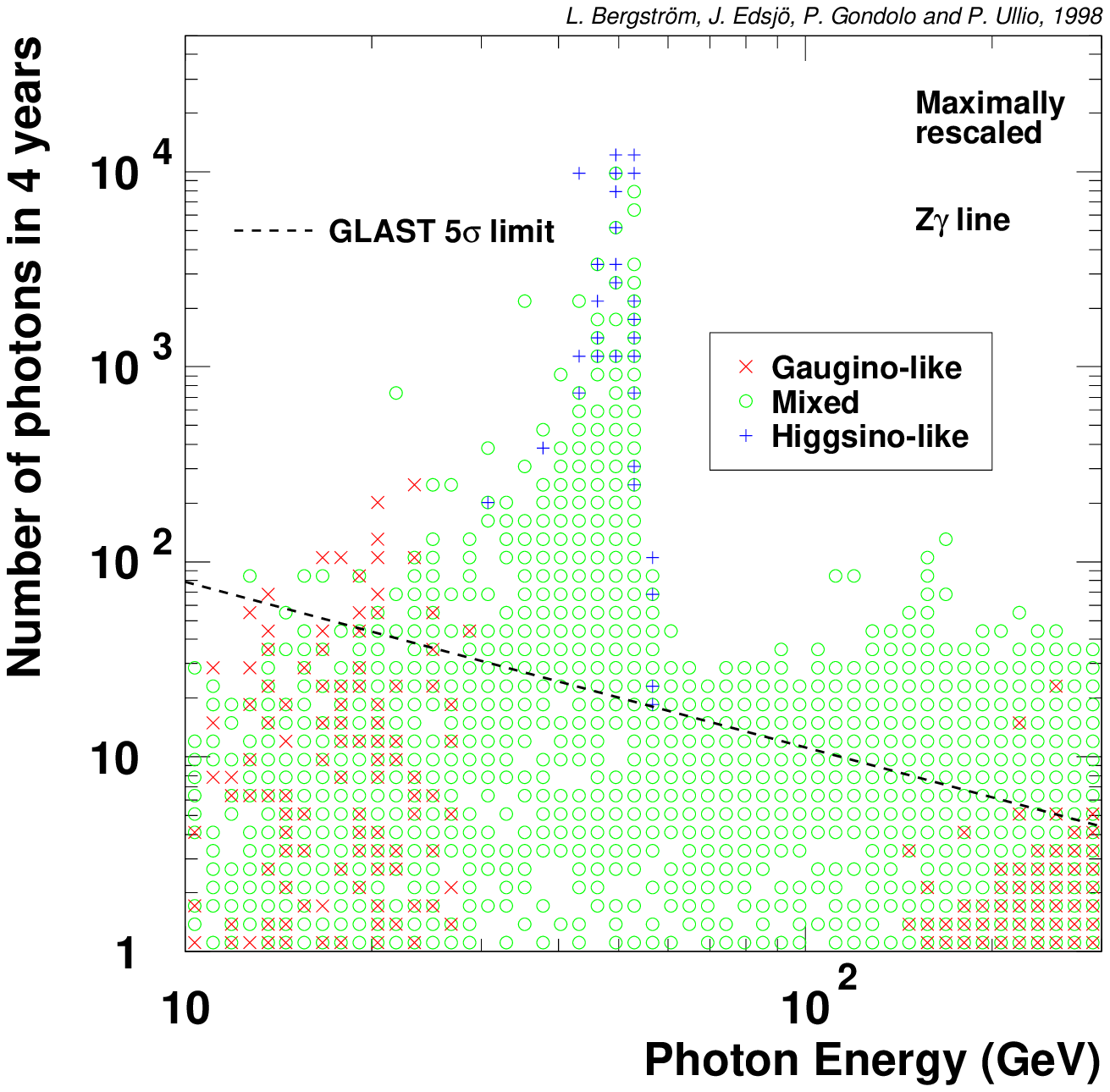,width=0.49\textwidth}} 
\caption{The number of expected photons in 0.84 sr towards 
$b=90^{\circ}$ collected in 4 years from (a) the $2\gamma$ and (b) the 
$Z\gamma$ final states.  The expected $5\sigma$ limit from the GLAST 
detector is also shown.}
\label{fig:lines}
\end{figure}

Taking into account the screening of the earth, the useful geometrical 
acceptance of GLAST towards a fixed point of the sky in a 0.84~sr cone 
is 0.21~m$^2$~sr~\cite{bloomprivate}; the energy resolution is assumed 
to be 1.5\%.  We display in Fig.~\ref{fig:lines} the number of 
expected $\gamma$s in 4 years of exposure time when the fluxes have 
been maximally rescaled according to 
Fig.~\ref{fig:resc}.  Also shown is the curve giving the minimum 
number of events needed to observe an effect at the $5\sigma$ level, 
where, in lack of data, we have assumed above 1~GeV a 2.7 power law
falloff for the diffuse $\gamma$-ray background and inferred its 
normalization from Ref.~\cite{sreekumar}.  As can be seen, a fair 
fraction of our set of supersymmetric models can be probed under these 
circumstances. Remember that the number of photons given in 
Fig.~\ref{fig:lines} is towards $b=90^{\circ}$ and, depending on halo 
profile, we expect more events towards the galactic centre, with a larger 
portion of the MSSM parameter space which might be probed.

\subsection{Diffuse neutrinos}

\begin{figure}
\centerline{\epsfig{file=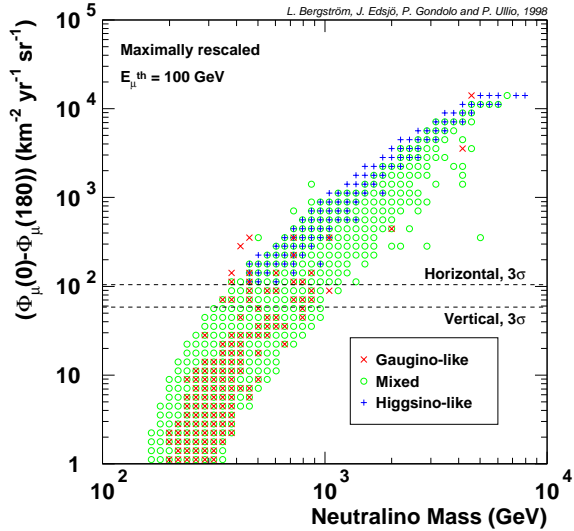,width=0.49\textwidth}}
\caption{The difference of the diffuse neutrino flux towards the 
galactic centre to that to the antigalactic centre. The fluxes are 
averaged over 2.5 sr which maximizes signal to noise and they are 
rescaled maximally as allowed by the antiproton and continuum 
$\gamma$ fluxes. The limits are for a neutrino telescope with an exposure 
of 10 km$^{2}$ yr.}
\label{fig:diffneutr}
\end{figure}

A possibility to get a detectable neutrino flux from WIMP annihilations 
which has rarely been considered in the literature is neutrinos from
annihilation in the galactic halo.  Of particular importance is
$\chi\chi\to W^+W^-$, since the $W$ bosons decay in 10 \% of the cases
directly to a muon plus a muon neutrino with a hard neutrino spectrum,
which may facilitate detection in neutrino telescopes.

This flux would scale in exactly the same way as the gamma flux in the 
presence of clumps and with future $\cal O$ (km$^{3}$) 
neutrino-telescopes, the diffuse neutrinos might prove more 
constraining than antiprotons and continuum $\gamma$s at high masses 
(several hundred GeV -- TeV region) where the rescaling can be high 
($f\delta > 10^{3}$).

The flux has been calculated in essentially the same way as for 
neutralino annihilation in the Sun/Earth \cite{beg} with the help of the Lund 
Monte Carlo {\sc Pythia} 6.115. The only difference is that some 
annihilation products will decay and produce neutrinos in the halo, 
whereas they are stopped before they decay in the Sun/Earth.

The neutrino-induced muon flux from neutralino annihilations in a 
smooth halo is about $10^{-8}$--1 km$^{-2}$ yr$^{-1}$ sr$^{-1}$ above 
100 GeV. Compare this with the atmospheric background of about 9500 
km$^{-2}$ yr$^{-1}$ sr$^{-1}$ vertically and 30000 km$^{-2}$ yr$^{-1}$ 
sr$^{-1}$ horizontally \cite{atm-nu} for this threshold.  To be able 
to distinguish the signal from the background we have to rescale the 
fluxes by the allowed clumpiness factor derived in the previous 
section and we also have to make use of the fact that the signal is 
enhanced towards the centre of the galaxy.

The best prospects are probably given by large-area neutrino telescopes 
with relatively high detection thresholds.  We can imagine measuring the 
flux in a solid angle $\Delta\Omega$ towards the galactic centre and 
compare with the flux in the same solid angle in the opposite 
direction. The limit we can put on the flux is at the $3\sigma$-level 
approximately given by
\begin{equation}
   \left[\Phi_{\mu}(0^{\circ}) - \Phi_{\mu}(180^{\circ})\right]_{\rm limit} 
   \simeq 3 \sqrt{\frac{\Phi_{bkg}}{{\cal E}\Delta\Omega}}.
\end{equation}
where $\cal E$ is the exposure.  For the modified isothermal sphere, 
it turns out the best limits are obtained with $\Delta\Omega = 
2.5$ sr, for which we obtain $\langle J(0^{\circ}) \rangle 
(\Delta\Omega) = 4.16$ and $\langle J(180^{\circ}) \rangle 
(\Delta\Omega) = 1.09$.  In Fig.~\ref{fig:diffneutr} we show the 
difference of the diffuse neutrino flux towards the galactic centre to 
that in the opposite direction for a muon energy threshold of 100 
GeV\@.  Also shown are the limits that can be reached with an exposure 
of 10 km$^{2}$ yr.  For different exposures, the limits change as the 
square root of the exposure.  If we increase the threshold from 100 
GeV, we can gain a small factor in sensitivity at higher masses, but 
lose at intermediate masses.

An ideal neutrino detector for this signal would view the galactic 
center through the center of the Earth (i.e.\ it should be at 29 
degrees latitude), since then the atmospheric background is minimal.  
The strength of the signal of course depends on the halo profile, but 
it is more likely that the halo profile is steeper towards the 
galactic centre than the isothermal sphere and hence the signal is 
even bigger then envisioned here. We might have to worry 
about other sources of high-energy neutrinos at the galactic centre 
(like neutrinos from the black hole believed to exist in the centre). 
These other sources can probably be removed by not looking at the very 
centre of the galaxy.

\subsection{Positrons}

\begin{figure}
\centerline{\epsfig{file=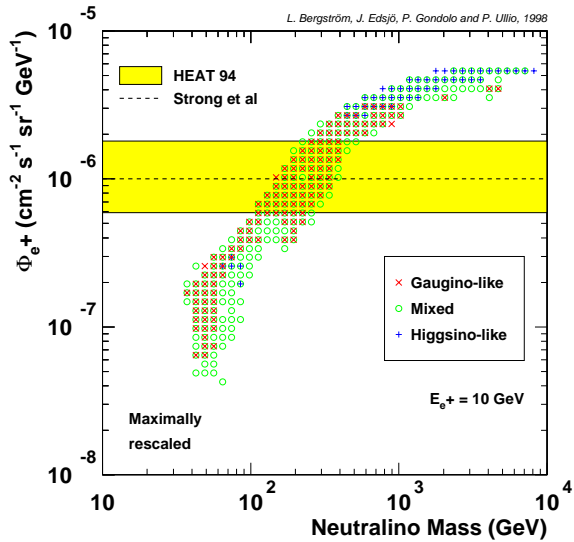,width=0.49\textwidth}}
\caption{The positron fluxes rescaled maximally as allowed by the 
antiproton and continuum $\gamma$ fluxes.  The 1994 HEAT measurement 
at 10 GeV \protect\cite{heat} is shown together with the Strong et 
al.\ \protect\cite{strong} prediction for the background at this energy.}
\label{fig:positrons}
\end{figure}

From neutralino annihilation in the halo we would also get a flux of 
positrons which might be detected by satellite \cite{ams} or 
high-flying balloon experiments \cite{heat}.  The propagation of 
positrons is a more difficult issue than for antiprotons since 
positrons are so easily deflected and destroyed.  We have calculated 
the positron fluxes using {\sc Pythia 6.115} \cite{pythia} and have 
used the propagation model in Ref.\ \cite{kamturner} with an energy 
dependent escape time (a more detailed investigation is in preparation 
\cite{bepos}).  In Fig.~\ref{fig:positrons} we show the positron 
fluxes versus the neutralino mass when they have been rescaled with the 
maximal $f\delta$ allowed by the antiproton and the continuum gamma 
fluxes.  We compare with the measurement by the HEAT experiment 
\cite{heat} at 10 GeV\@. Also shown is the prediction of the 
background at this energy as given in Ref.\ \cite{strong}.
It would seem that the positrons put more stringent bounds on 
$f\delta$ than the antiprotons and continuum $\gamma$s at higher 
masses.  The positron fluxes are however even more uncertain than the 
antiproton fluxes and can easily be wrong by a factor of 10.  Hence we 
can't use the positrons to constrain $f\delta$ further, but we see 
that we might be able to get measurable fluxes.

\subsection{Direct detection}

For the direct rates, we have used the procedures in Ref.~\cite{bg}. 
Since these rates only depend on the local halo density at present, 
they will as expected not put any severe constraints on the clumpiness of
 the 
halo as a whole. They will of course be much enhanced if we happen to be
inside a clump at present.
As with the neutrinos from neutralino annihilation in 
the Sun/Earth we can however have correlations between these signals 
and the signals giving high $\bar{p}$ or $\gamma$ fluxes. These correlations
are not very strong, however.

\subsection{Neutralino annihilation in the Earth/Sun}

Neutrinos are  produced in the annihilations through the decays
of quarks, leptons and gauge bosons produced in the primary annihilation
process. During the several billion years the Earth and Sun have 
existed, there may have been a substantial accumulation of neutralinos
due to capture, i.e.\ scattering and subsequent gravitational 
trapping. 

The fluxes of neutrino-induced muons from neutralino annihilation in 
the Earth/Sun are mostly determined by the capture rates, which in turn 
depend on the local halo density.  They are thus insensitive to 
different halo profiles; if the halo is clumped, there can however be 
fluctuations in the capture rates by time, but on the average we will 
capture the same amount of neutralinos as without clumps.  The amount 
of these fluctuations in capture rate and consequently in 
annihilation rate depends on the time between encounters, the size of 
the clumps, $r_{cl}$, and their overdensity $\delta$. For the 
small-clumps scenario these fluctuations are expected to be small.

Since these rates do not depend strongly on the clumping, they will not 
put better constraints on the clumpiness than the antiproton fluxes or 
the continuum gammas.


\section{Conclusions}

To conclude, we have found that the limits on antiproton and 
continuum $\gamma$ fluxes already constrain in 
a non-trivial way the clumpiness 
of the Milky Way dark matter halo (if made of neutralinos).
 The general pattern is that at lower 
masses the $\bar{p}$ 
flux puts the best limits, but at higher masses the continuum 
$\gamma$ flux is better.  Within the MSSM, the allowed 
clumpiness is less than $f\delta \simeq 10^{9}$ for all neutralino 
masses (assuming that the neutralinos make up most of the dark matter of 
our galaxy). 

We have also investigated what the detection prospects would be for 
other dark matter searches in this clumpy scenario, where the maximal 
rescaling is given by the limits on the antiproton and continuum 
$\gamma$ fluxes.  We have found that the fluxes of monochromatic 
$\gamma$-lines from halo annihilation into the $Z\gamma$ and $2\gamma$ 
final states can be enhanced enough to be seen by upcoming experiments 
like GLAST\@. We have also found that the flux of positrons from 
neutralino annihilation in the halo gets high enough to even exceed 
the current limits from the HEAT experiment.  The uncertainties for 
the positron flux are particularly large, however, and at present we 
can merely conclude that it is possible to obtain measurable fluxes of 
positrons.  The rarely-mentioned diffuse neutrino flux from neutralino 
annihilation in the halo can, for heavy neutralinos, be boosted enough 
to show a detectable difference in flux towards the galactic centre 
and the galactic anticentre.

It is reassuring that new detectors, like GLAST for gamma rays and AMS
for antiprotons, will obtain more stringent bounds on the clumpiness
of the Milky Way halo.  And, of course, finding evidence for
neutralino annihilations in the halo would be one of the most
important scientific discoveries of our time.

\section*{Acknowledgments}

This work was supported, in part, by the Human Capital and Mobility 
Program of the European Economic Community under contract No.\  
CHRX-CT93-0120 (DG 12 COMA).  L.B. was supported by the Swedish 
Natural Science Research Council (NFR).  J.E. would like to thank the 
hospitality of Max Planck Institut f\"ur Physik in M{\"u}nchen where 
parts of this work were completed.  We thank E.~Bloom for providing 
information of GLAST. This work was supported with computing resources 
by the Swedish Council for High Performance Computing (HPDR) and 
Parallelldatorcentrum (PDC), Royal Institute of Technology.

\end{document}